# Tunable Photoluminescence of Monolayer MoS$_2$ via Chemical Doping

Shinichiro Mouri[1,*], Yuhei Miyauchi[1,2], and Kazunari Matsuda[1,+]

[1]Institute of Advanced Energy, Kyoto University, Uji, Kyoto 611-0011, Japan

[2]Japan Science and Technology Agency, PRESTO, 4-1-8 Honcho Kawaguchi, Saitama 332-0012, Japan

**Abstract**

We demonstrate the tunability of the photoluminescence (PL) properties of monolayer (1L)-MoS$_2$ via chemical doping. The PL intensity of 1L-MoS$_2$ was drastically enhanced by the adsorption of *p*-type dopants with high electron affinity, but reduced by the adsorption of *n*-type dopants. This PL modulation results from switching between exciton PL and trion PL depending on carrier density in 1L-MoS$_2$. Achievement of the extraction and injection of carriers in 1L-MoS$_2$ by this solution-based chemical doping method enables convenient control of optical and electrical properties of atomically thin MoS$_2$.

Atomically thin transition-metal dichalcogenides (TMDs) have attracted a great deal of attention from the viewpoints of fundamental physics and various applications.[1–4] The thin-layered TMDs, as novel two-dimensional materials, undergo remarkable changes in their electronic structures depending on the number of layers: from indirect-band-gap bulk semiconductors to direct-band-gap monolayer semiconductors. Monolayer $MoS_2$ (1L-$MoS_2$) and its analogues ($MoSe_2$, $WS_2$, $WSe_2$, etc.) have potential applications as novel two-dimensional direct-band-gap semiconductors[1–9] in various opto-electronic devices,[3,4] such as low-electricity-consumption transistors,[10,11] phototransistors,[12] light-emitting devices,[13] and solar cells.[14,15] Moreover, 1L-$MoS_2$ is emerging as a novel platform for the study of "spintronics" or "valleytronics" because of its broken inversion symmetry, which gives rise to the coupling of spins and the valley degree of freedom in the momentum space.[16–19]

Optically generated electron–hole pairs in 1L-$MoS_2$ form stable exciton states even at room temperature[1,2] because of the extremely large Coulomb interactions in atomically thin two-dimensional materials. The stable exciton plays an important role in the optical properties of 1L-$MoS_2$. Control of the carrier density is one effective method to modulate the optical properties of monolayer TMDs.[20–23] The interplay between the exciton and charge carrier gives rise to the formation of a many-body bound state such as a charged exciton (trion), which provides additional pathways for controlling the optical properties of 1L-$MoS_2$.[20] The application of gate bias voltages using field-effect transistor (FET) devices[20–22] and gas physisorption have been utilized for carrier doping;[23] however, the complicated FET structure or the precise control of gas regulation is required to achieve doping by these methods, which could limit the fundamental study and application of carrier-doped 1L-$MoS_2$. An alternative doping method for controlling the optical properties in 1L-$MoS_2$ is strongly desired. Chemical doping is known to be an effective and convenient method to modify the carrier density of monolayer materials, including graphene[24,25] and single-walled carbon nanotubes.[26–29] The charge transfer between the dopant molecules and the monolayer materials is expected to induce a shift in the Fermi level and to enable large modulations of the optical and electrical properties of monolayer materials.

In this letter, we demonstrate that the photoluminescence (PL) properties of 1L-MoS$_2$ can be tuned via a solution-based chemical doping technique. The PL intensity of 1L-MoS$_2$ is drastically enhanced when *p*-type dopants cover its surface. This enhancement is understood as a consequence of switching the dominant PL process from the recombination of the negative trion to the recombination of the exciton under extraction of residual electrons in as-prepared 1L-MoS$_2$. In contrast, the PL intensity is reduced when 1L-MoS$_2$ is covered with *n*-type dopants, which is due to the suppression of exciton PL by the injection of excess electrons. On the basis of these results, we confirmed that bidirectional control of the Fermi level of 1L-MoS$_2$ by chemical doping is possible and that this approach provides a great advantage in controlling the optical and electrical properties of thin-layered MoS$_2$.

MoS$_2$ flakes were mechanically exfoliated from bulk crystals onto 300-nm-thick SiO$_2$/Si substrates. The optical images and Raman and PL spectra enable us to identify the number of layers[1,2,30–32], as described below. Micro-PL measurements for chemical doping of MoS$_2$ at room temperature were conducted using a home-built optical setup. A solid-state laser diode (2.33 eV) was used for the Raman and PL measurements. The objective lens (NA = 0.8) was used, and the focused laser spot size on the sample was less than 1 μm$^2$. Typical excitation power was maintained at less than 20 μW to avoid any heating and nonlinear optical effects. The PL signal was detected using a liquid-nitrogen-cooled CCD camera through the monochromator, and the typical exposure time was 120 seconds. We used 2,3,5,6-tetrafluoro-7,7,8,8-tetracyanoquinodimethane (F$_4$TCNQ) and 7,7,8,8-tetracyanoquinodimethane (TCNQ) as *p*-type chemical dopants and nicotinamide adenine dinucleotide (NADH) as *n*-type dopants. The solution-based chemical doping was performed using a drop-cast method: an approximately 10-μl droplet of the dopant solution was pipetted onto the SiO$_2$/Si substrate area of 1 cm$^2$ where the flakes of 1L-MoS$_2$ were prepared. The concentration of the dopant solution was 0.02 μmol/ml, which corresponds to a dopant density on the 1L-MoS$_2$ of approximately 1/nm$^2$ after one doping step. All optical measurements were performed after evaporation of the solvent solution. We performed sequential doping by repeating this procedure.

Figure 1a shows the typical optical images of the as-prepared monolayer (1L), bilayer (2L), and trilayer (3L) MoS$_2$. The Raman and PL spectra of these samples are shown in Fig. 1b and Fig. 1c, respectively. The frequency difference between out-of-plane (A$_{1g}$: ~404 cm$^{-1}$) and in-plane (E$^1_{2g}$: ~385 cm$^{-1}$) phonon modes in the Raman spectra depends on the number of layers of MoS$_2$ in the case of less than 10 monolayers and increases with an increase in the number of layers. The frequency differences of the two modes at 19.2 cm$^{-1}$ in 1L-MoS$_2$, 22.3 cm$^{-1}$ in 2L-MoS$_2$, and 24.4 cm$^{-1}$ in 3L-MoS$_2$ are consistent with previously reported results.[31,32]

Figure 1c shows the PL spectra of 1L-, 2L-, and 3L-MoS$_2$. Two major peaks at approximately 1.85 eV (A) and 2.05 eV (B) appear in the PL spectra of MoS$_2$. The A and B PL peaks are associated with the direct gap transition at K (K′) point,[1,2] and the energy difference between the A and B peaks corresponds to the valence-band splitting due to the strong spin–orbital interaction. The weak PL peaks (I) associated with the indirect band-gap transition are observed in the PL spectra of 3L- and 2L-MoS$_2$ and disappear in the spectrum of 1L-MoS$_2$ because of the transition of indirect to direct band-gap energy structures from 2L- to 1L-MoS$_2$.[2] The PL intensity of the A peak increases with a decrease in the number of layers. These PL features are consistent with those reported in previous studies.[1,2]

Figure 2a shows PL spectra of 1L-MoS$_2$ with and without F$_4$TCNQ doping. The PL intensity of the A peak in the as-prepared 1L-MoS$_2$ is very weak, whereas the PL intensity is drastically enhanced with F$_4$TCNQ doping. The PL intensity increases step by step with increases in the F$_4$TCNQ doping steps, as shown in Fig. 2b, and is almost saturated in the case of more than 10 doping steps, as shown in the Fig. 2d. The total integrated intensity of the peak A ($I_{total}$: black circles in Fig. 2d) is approximately three times greater than that of as-prepared 1L-MoS$_2$.

The PL spectral shape of 1L-MoS$_2$ also changed upon F$_4$TCNQ doping. Figure 2c shows the PL spectra normalized by their peak height. The spectral shape of peak A becomes sharp after F$_4$TCNQ doping. The peak energy of the PL spectra in 1L-MoS$_2$ is blue-shifted with an increase in the number of doping steps and plateaus at approximately 1.88 eV, as shown in the inset of Fig. 2b. To understand these spectral changes, we consider the contributions of the exciton (X) and the trion ($X^-$) in PL peak A.[20] Peak A can be decomposed into the exciton (X; ~1.88 eV; red line) peak and the negative trion ($X^-$; ~1.84 eV; blue line) peak under the assumption that both are Lorentzian peaks, as shown in Fig. 2c. The PL spectra of 1L-MoS$_2$ are well reproduced by the sum of these two peak components and the B exciton peak (Supporting Information S1). This analysis clearly suggests that the PL spectral weight of the negative trion peak ($X^-$) at ~1.84 eV is greater than that of the exciton peak (X) at ~1.88 eV in the as-prepared 1L-MoS$_2$. This experimental fact is consistent with previously reported results, where the trion ($X^-$) recombination has been reported to be dominant in the as-prepared 1L-MoS$_2$ on account of unintentional heavy electron (*n*-type) doping.[20] The PL spectra of 1L-MoS$_2$ after F$_4$TCNQ doping are dominated by the exciton peak (X) at ~1.88 eV, which strongly suggests that the excitons can recombine without forming trions because of the decrease in the number of excess carriers in 1L-MoS$_2$.

Figure 2d shows the integrated PL intensity of excitons ($I_x$: red triangles) and trions ($I_x^-$: blue triangles) as a function of the number of doping steps. The integrated PL intensity of exciton $I_x$ monotonically increases with an increase in the number of doping steps and saturates when the number of doping steps exceeds 10. In contrast, the integrated PL intensity of trion $I_x$- exhibits a slight decrease. As a result, the total PL intensity of peak A in the spectrum of 1L-MoS$_2$ ($I_{total}$: black circles) drastically increases, primarily because of the enhancement of the exciton PL intensity. This behavior is similar to the PL spectral change induced by FET gate doping.[20]

We also investigated the change in the PL properties of 1L-MoS$_2$ induced by a different *p*-type dopant and an *n*-type dopant (i.e., TCNQ and NADH, respectively), as shown in Figs. 3a and 3b, respectively. A drastic enhancement in PL intensity, which is similar to that induced by F$_4$TCNQ doping, is also observed after TCNQ doping. On the basis of the PL spectral shape analysis, we confirmed that the spectrum of TCNQ-doped 1L-MoS$_2$ is also dominated by the exciton PL (see Supporting Information S2). In contrast, the PL intensity of 1L-MoS$_2$ is suppressed by the NADH *n*-type dopant molecules, as shown in Fig. 3b. The PL intensity suppression arises primarily from the decrease in the exciton PL intensity, whereas the trion PL weight is increased (see Supporting Information S2). These results suggest that the electrons from the *n*-type dopant (NADH) are additionally doped into 1L-MoS$_2$. The PL changes in 1L-MoS$_2$ show opposite behaviors in the cases of *p*-type dopants (F$_4$TCNQ and TCNQ) and *n*-type dopants (NADH).

These experimental results provide strong evidence with respect to the chemical doping mechanism for 1L-MoS$_2$. As shown in Fig. 3c, the reported flat band potential of few-layer MoS$_2$ is −0.13 eV (vs. SHE),[33] which is smaller than the reduction potentials of 0.84 eV (vs. SHE) for F$_4$TCNQ and 0.46 eV (vs. SHE) for TCNQ.[34] Therefore, these *p*-type dopants function as electron acceptors for 1L-MoS$_2$. In contrast, the *n*-type dopant (NADH) functions as an electron donor due to its low oxidation potential (−0.32 eV[35] vs. SHE). The difference in chemical potential between 1L-MoS$_2$ and *p*-type (*n*-type) dopants induces the electron extraction (injection).

To further understand the behavior of the PL intensity of the exciton $I_x$ and trion $I_{x^-}$, we discuss the excited dynamics within the framework of a three-level model that includes a trion, an exciton, and the ground state, as shown in Fig. 4a. The term *G* represents the generation rate of A excitons. The decay rates of the excitons and trions are denoted as $\Gamma_{ex}$ and $\Gamma_{tr}$, respectively. Assuming that the rate of adsorption of F$_4$TCNQ molecules onto 1L-MoS$_2$ obeys Langmuir's law, the rate of formation of trions from the excitons $k_{tr}(n)$ after *n*-th doping steps can be described as

$$k_{tr}(n) = k_{tr}(0)(1 - s\theta) \quad , \tag{1}$$

where *s* reflects the charge transfer ability from a F$_4$TCNQ molecule to 1L-MoS$_2$ on a per molecule basis and $\theta$ is the fractional coverage of the surface. $\theta$ is given as $\theta = \alpha n \delta/(1+ \alpha n \delta)$ where $\delta$ (i.e., 0.02 µmol/ml) is the F$_4$TCNQ concentration, and $\alpha$ is a parameter that reflect the adsorption probability. The dependence of the PL intensity of the exciton (X) and trion (X$^-$) on the number of doping steps can be evaluated by solving the rate equations within the framework of a three-level model in the steady-state condition. The calculated PL intensities $I_x$ and $I_{x^-}$ of the exciton and trion are indicated by the red and blue solid lines, respectively, in Fig. 2d (see Supporting Information S3); these calculated intensities well reproduce the experimental results. The parameters used here, i.e., $\Gamma_{ex} = 0.002$ ps$^{-1}$, $\Gamma_{tr} = 0.02$ ps$^{-1}$, and $k_{tr}(0) = 0.5$ ps$^{-1}$, are consistent with the previously reported values obtained from transient absorption measurements.[36] The very fast trion formation process ($k_{tr}(0) = 0.5$ ps$^{-1}$) is dominant in the exciton decay dynamics in the as-prepared 1L-MoS$_2$ due to the unintentionally heavily doped electrons. The trion formation rate is strongly reduced from $k_{tr}(0) = 0.5$ ps$^{-1}$ to $k_{tr}(n) = 0.02$ ps$^{-1}$ above the doping step of 10, where the enhancement of the total PL intensity $I_{total}$ is saturated. This mechanism causes the strong enhancement of the exciton PL in F$_4$TCNQ-doped 1L-MoS$_2$.

Figure 4b shows the doping step dependence of the trion PL spectral weight, which is defined as $I_{x^-}/I_{total}$. The trion PL spectral weight decreases gradually from 0.65 to 0.15 and saturates with an increase in the number of doping steps, which suggests that the PL spectrum of 1L-MoS$_2$ is dominated by the exciton peak in the saturation region. Assuming the mass action law is valid (see Supporting Information S4), the electron density $n_{el}$ can be evaluated from the trion PL spectral weight, and the calculated electron density $n_{el}$ as a function of the number of F$_4$TCNQ doping steps is shown in the inset of Fig. 4b. The estimated electron densities of 1L-MoS$_2$ before and after doping in the saturation region are ~6 × 10$^{13}$ and ~2 × 10$^{12}$ cm$^{-2}$, respectively. Given that the typical molecular cross-section of F$_4$TCNQ is ~0.1 nm$^2$,[37] the limitation of the possible density of F$_4$TCNQ molecules on a 1L-MoS$_2$ surface is ~10 /nm$^2$. This value is consistent with the density of dropped F$_4$TCNQ after 10 doping steps if the dopant density is assumed to be ~1/nm$^2$ in each doping step. On the basis of this estimation, the surface of 1L-MoS$_2$ is fully covered by the F$_4$TCNQ molecules at the saturation region, which also suggests that the saturation of the PL intensity enhancement is caused by the limitation of the surface adsorption of dopant molecules.

In summary, we demonstrated the tunable PL properties of 1L-MoS$_2$ using a molecular chemical doping technique. The PL intensity of 1L-MoS$_2$ was drastically enhanced by the adsorption of $p$-type dopants (F$_4$TCNQ and TCNQ). This intensity enhancement was explained by the switching of the dominant PL process from the recombination of negative trions to the recombination of excitons through extraction of the unintentionally highly doped electrons. Moreover, the PL intensity was reduced by the adsorption of *n*-type dopants (NADH), which we attributed to the suppression of exciton PL through injection of the excess electrons. Our findings suggest that both the extraction and the injection of electrons in 1L-MoS$_2$ can be realized via the solution-based chemical doping technique, which provides a strong advantage in tuning the optical and electrical properties of atomically thin TMDs without the use of device structures.

ACKNOWLEDGMENTS

This study was supported by a Grant-in-Aid for Scientific Research from MEXT of Japan (Nos. 25400324, 60516037, 40311435, 23340085 and 24681031) and by PREST from JST.

# REFERENCES

*E-mail: iguchan@iae.kyoto-u.ac.jp

+E-mail: matsuda@iae.kyoto-u.ac.jp

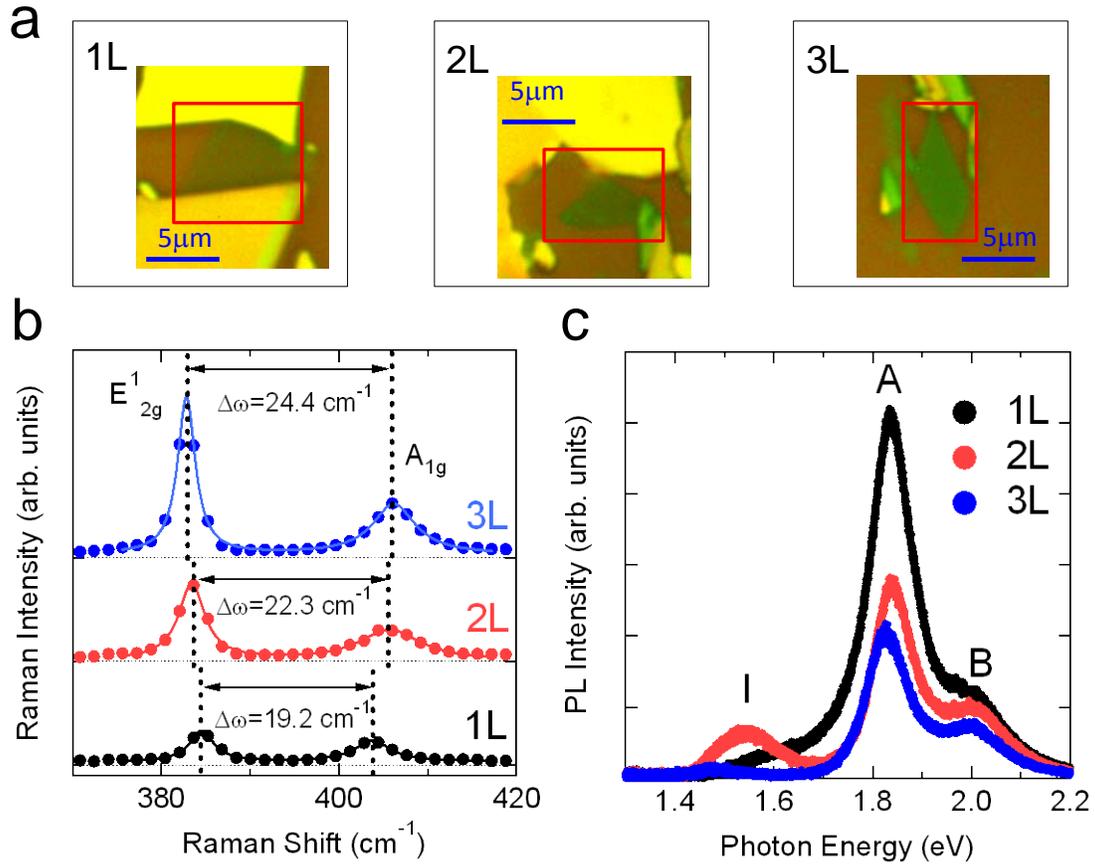

**Figure 1.** (a) Optical images of as-prepared monolayer (1L)-, bilayer (2L)-, and trilayer (3L)-$MoS_2$ on $SiO_2$/Si substrates. (b) Raman spectra of the as-prepared 1L-, 2L-, and 3L-$MoS_2$ measured at room temperature. (c) PL spectra of the as-prepared 1L-, 2L-, and 3L-$MoS_2$. The PL peak due to the indirect band-gap transition is denoted as I, and those due to the direct band-gap transition are denoted as peaks A and B.

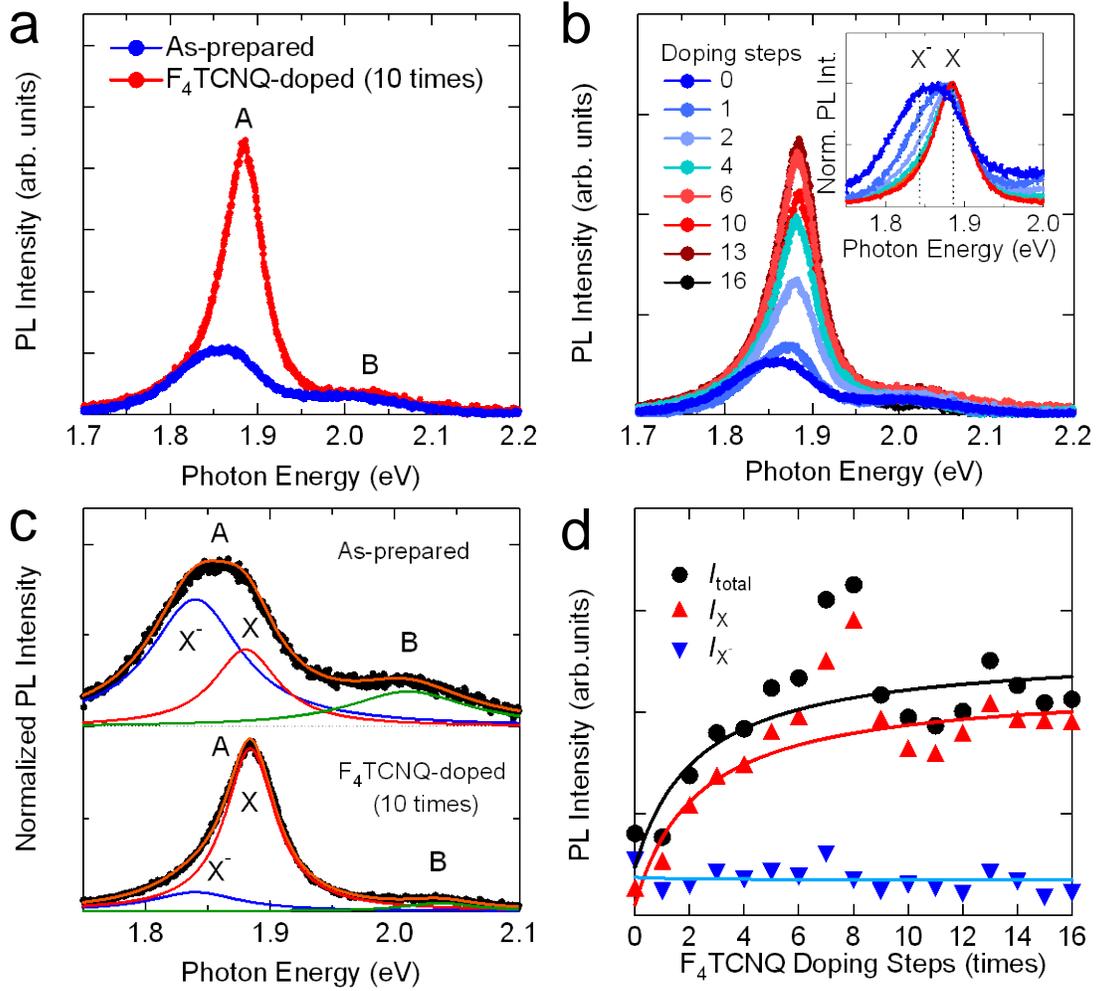

**Figure 2.** (a) PL spectra of 1L-MoS$_2$ before and after F$_4$TCNQ doping. (b) PL spectra of 1L-MoS$_2$ obtained at each doping step (0, 1, 2, 4, 6, 10, 13, and 16 steps). The inset shows the normalized PL spectra of 1L-MoS$_2$ at each doping step. (c) Analysis of the PL spectral shapes for as-prepared and F$_4$TCNQ-doped 1L-MoS$_2$. The A peaks in the PL spectra were reproduced by assuming two peaks with Lorentzian functions, corresponding to the trion (X$^-$) and the exciton (X) peaks, were overlapped. (d) Integrated PL intensity of the negative trion $I_{X^-}$, exciton $I_X$, and the sum ($I_{total}$) of $I_X$ and $I_{X^-}$ as functions of the number of F$_4$TCNQ doping steps. Solid lines show the calculated PL intensity curves calculated by solving the rate equations in the three-level model.

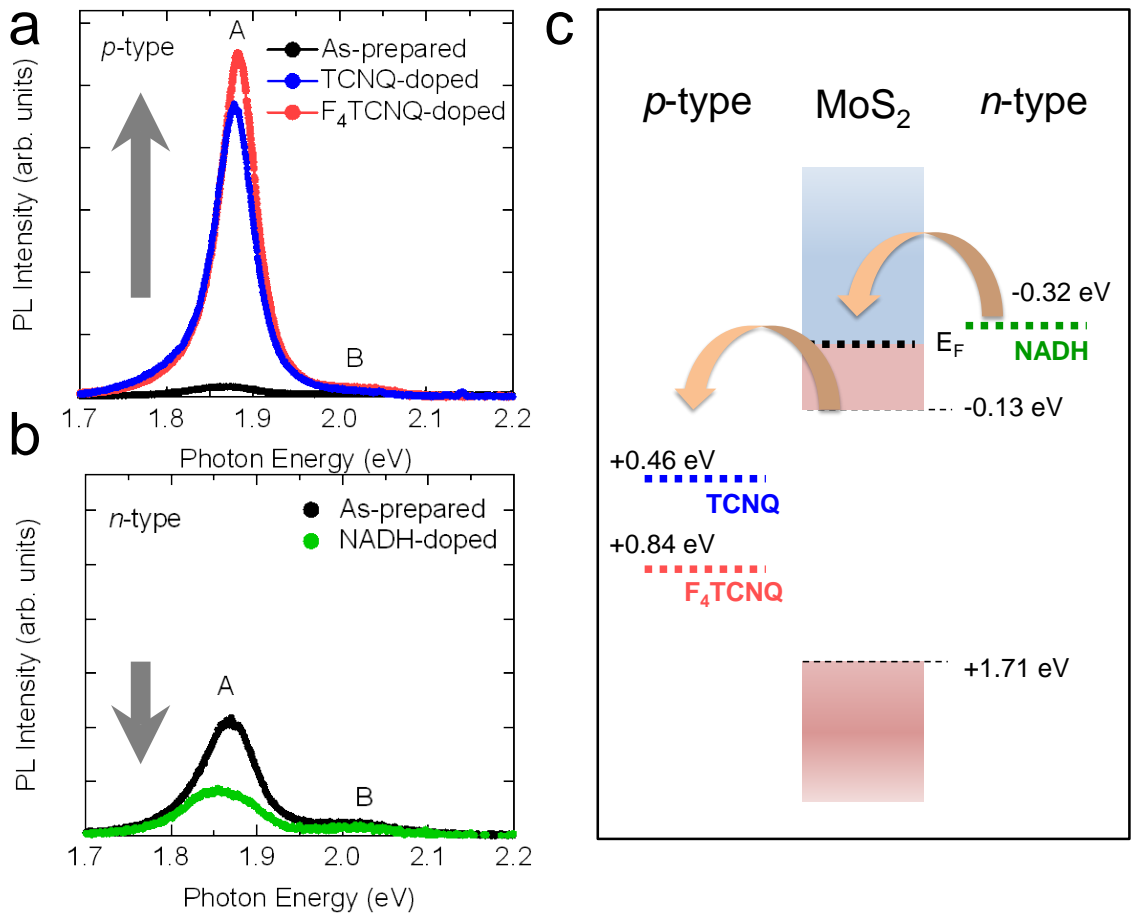

**Figure 3.** (a) PL spectra of 1L-MoS$_2$ before and after being doped with *p*-type molecules (TCNQ and F$_4$TCNQ). (b) PL spectra of 1L-MoS$_2$ before and after being doped with an *n*-type dopant (NADH). (c) Schematic of relative potentials (vs. SHE) of 1L-MoS$_2$ and *n*- and *p*-type dopants.

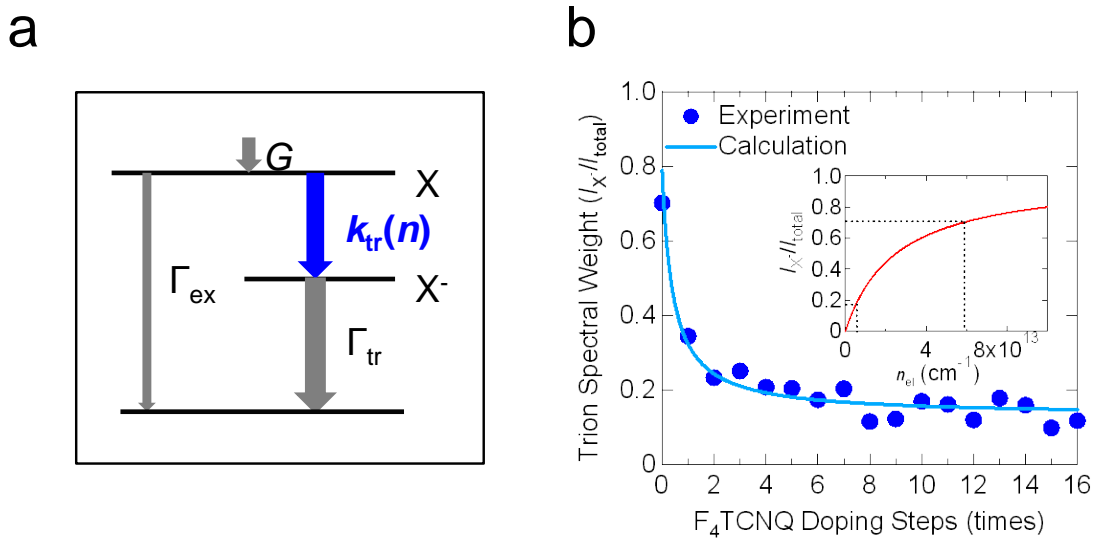

**Figure 4.** (a) Three-level energy diagram including the exciton (X), the trion (X$^-$), and the ground state. (b) PL intensity weight of the trion, which is defined as $I_{x^-}/I_{total}$, as a function of the number of F$_4$TCNQ doping steps. The calculated PL intensity weight of the trion is indicated by the solid curve. The inset shows the electron density ($n_{el}$) calculated by the law of mass action (see Supporting Information S4).

Supporting information for

# Tunable Photoluminescence of Monolayer MoS$_2$ via Chemical Doping

Shinichiro Mouri[1,*], Yuhei Miyauchi[1,2], and Kazunari Matsuda[1,+]

[1]Institute of Advanced Energy, Kyoto University, Uji, Kyoto 611-0011, Japan

[2]Japan Science and Technology Agency, PRESTO, 4-1-8 Honcho Kawaguchi, Saitama 332-0012, Japan

**1. Series of data of decomposed results of PL spectra**

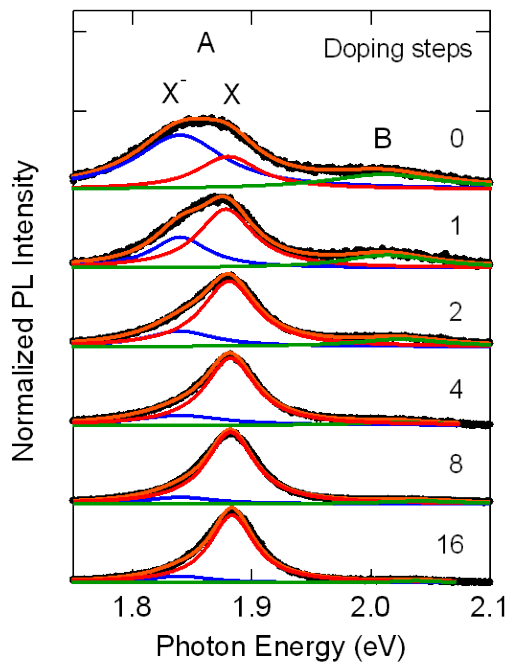

Fig. S1 Series of decomposed PL spectra of F$_4$TCNQ-doped 1L-MoS$_2$

Figure S1 shows the decomposed PL spectra of $F_4$TCNQ-doped 1L-$MoS_2$ at each doping step. All spectra are normalized by the peak height. The PL spectra can be decomposed into three components: A exciton PL (X; ~1.88 eV; red line), A trion PL ($X^-$; ~1.84 eV; blue line), and B exciton PL (B; 2.05 eV; green line), as described in the main text. Although the shape of the PL peak differs in each 1L-$MoS_2$, reflecting the doped electron density, all PL spectra are well reproduced (orange lines) by the sum of three peak components.

## 2. PL spectral change of 1L-$MoS_2$ with $p$-type (TCNQ) and $n$-type (NADH) dopants.

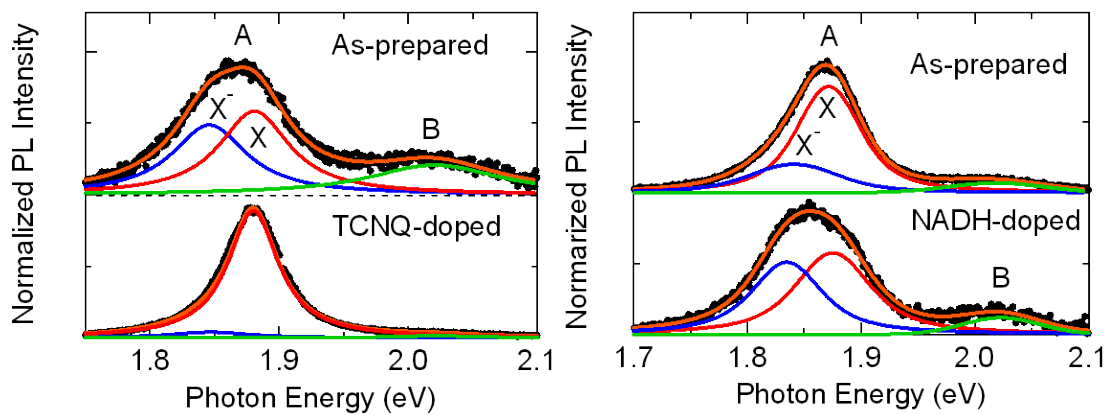

Fig. S2 PL spectra of 1L-$MoS_2$ before and after chemical doping with $p$-type (TCNQ) and $n$-type dopants (NADH).

Figure S2 shows the decomposed PL spectra of 1L-$MoS_2$ before and after chemical doping with $p$-type (TCNQ) and $n$-type dopants (NADH). These PL spectra also can be decomposed into three components: A exciton PL (X; ~1.88 eV; red line), A trion PL ($X^-$; ~1.84 eV; blue line), and B exciton PL (B; 2.05 eV; green line). The shape of the PL spectra of the as-prepared 1L-$MoS_2$ differs depending on the sample's initially doped electron density. We prepared less-electron-doped 1L-$MoS_2$ for the NADH doping experiment to confirm the electron injection effect.

## 3. Analysis of PL intensity of excitons and trions in 1L-MoS$_2$ with chemical doping within the framework of a three-level model.

The rate equations for the population of exciton $N_X$ and trion $N_{X^-}$ can be expressed as

$$\frac{dN_X}{dt} = G - \{\Gamma_{ex} + k_{tr}(n)\} N_X, \qquad (S1)$$

$$\frac{dN_{X^-}}{dt} = k_{tr}(n) N_X - \Gamma_d N_{X^-}, \qquad (S2)$$

where $n$ is the number of doping steps, $k_{tr}(n)$ is the formation rate of the trion from the exciton, and $G$ is the optical generation rate of excitons. The populations of excitons and trions derived from the steady-state solutions of these equations are expressed as

$$N_X(n) = \frac{G}{\Gamma_{ex} + k_{tr}(n)}, \qquad (S3)$$

$$N_{X^-}(n) = \frac{k_{tr}(n)}{\Gamma_{tr}} \frac{G}{\Gamma_{ex} + k_{tr}(n)}, \qquad (S4)$$

where the decay rates of the exciton and trion are denoted as $\Gamma_{ex}$ and $\Gamma_{tr}$, respectively. The PL intensity of the exciton ($I_X$) and trion ($I_X^-$) can be expressed on the basis of the relationship where the PL intensity is proportional to the exciton (trion) populations as follows:

$$I_X(n) = \frac{G\gamma_{ex}}{\Gamma_{ex} + k_{tr}(n)}, \qquad (S5)$$

$$I_{X^-}(n) = \frac{k_{tr}(n)}{\Gamma_{tr}} \frac{G\gamma_{tr}}{\Gamma_{ex} + k_{tr}(n)}, \qquad (S6)$$

where $\gamma_{ex}$ and $\gamma_{tr}$ express the radiative decay rate of the exciton and trion, respectively. The parameters $\Gamma_{ex} = 0.002$ ps$^{-1}$, $\Gamma_{tr} = 0.02$ ps$^{-1}$, and $k_{tr}(0) = 0.5$ ps$^{-1}$ in this analysis were based on the previously reported values obtained from transient absorption measurements.[S1] The best-fit parameters of $G\gamma_{ex}$ and $G\gamma_{tr}$ to reproduce the experimental results shown in Fig. 2d are 10 and 1.5, respectively. In the condition studied here (i.e., $k_{tr} \gg \Gamma_{ex}$), the PL intensity of the exciton ($I_x$) and trion ($I_x$-) can be approximately expressed as

$$I_X(n) \approx \frac{G\gamma_{ex}}{k_{tr}(n)}, \tag{S7}$$

$$I_{X^-}(n) \approx \frac{G\gamma_{ex}}{\Gamma_{tr}}, \tag{S8}$$

As shown in Fig. 2d, the experimental behavior of the exciton PL intensity $I_x$ strongly depends on the extent of F$_4$TCNQ doping, whereas the trion PL intensity $I_x$- is insensitive to the level F$_4$TCNQ doping; these results can be explained by eqs. (S7) and (S8).

4. **Mass action model**

The mass action law associated with the trions[S2] is used to evaluate the doped electron density in 1L-MoS$_2$. In this scheme, the following relation is obtained:

$$\frac{N_X n_{el}}{N_{X^-}} = \left(\frac{4 m_X m_e}{\pi \hbar^2 m_{X^-}}\right) k_B T \exp\left(-\frac{E_b}{k_B T}\right), \tag{S9}$$

where $T$ is the temperature, $k_B$ is the Boltzmann constant, $E_b$ is the trion binding energy (~20 meV),[S3] and $m_e$ (0.35$m_0$) and $m_h$ (0.45$m_0$) are the effective mass of electrons and holes,[S4] respectively, where $m_0$ is the mass of a free electron. The effective masses of an exciton $m_X$ and a trion $m_{X^-}$ can be calculated as 0.8$m_0$ and 1.15$m_0$, respectively. Using these parameters, the trion PL intensity weight is expressed as

$$\frac{I_{X^-}}{I_{total}} = \frac{\frac{\gamma_{tr}}{\gamma_{ex}}\frac{N_{X^-}}{N_X}}{1+\frac{\gamma_{tr}}{\gamma_{ex}}\frac{N_{X^-}}{N_X}} \approx \frac{4\times 10^{-14} n_{el}}{1+4\times 10^{-14} n_{el}}. \tag{S10}$$

The trion PL intensity weight calculated from eq. (S10) is shown in the inset of Fig. 4b.